\newfont{\abfont}{amr10 scaled\magstep1}
\newfont{\aufont}{amr10 scaled\magstep2}
\newfont{\tifont}{ambx10 scaled\magstep3}

\def\@maketitle{\newpage   
 \null
 \vspace*{-1\headsep}      
 \vspace*{-1\headheight}
 \vspace*{-24pt}
 \begin{flushright}{\large         
   { \preprintno} \\ \@date}
 \end{flushright}
 \vskip \headsep           
 \vskip \headheight
 \bigskip
 \begin{center}            
   {\LARGE\tifont \@title \par}
   \vskip 2em
   {\large\aufont
     \lineskip .5em
     \begin{tabular}[t]{c}\@author
     \end{tabular}\par}
   \vskip 1em
 \end{center}
 \par
 \vskip 1.5em}

\newcommand{\preprintno}{preprint number here}   

\def\abstract{\if@twocolumn
\section*{Abstract}
\else \large\abfont                             
\begin{center}
{\bf Abstract\vspace{-.5em}\vspace{0pt}}
\end{center}
\quotation
\fi}
\def\endabstract{\if@twocolumn\else\endquotation\fi}

\def\appendix{\par
    \setcounter{section}{0}
    \setcounter{subsection}{0}
    \renewcommand{\theequation}{\Alph{section}.\arabic{equation}}
    \setcounter{equation}{0}
}

\@addtoreset{equation}{section}
\def\theequation{\arabic{section}.\arabic{equation}}

\def\ps@columns{%
 \if@twocolumn
  \let\@mkboth\@gobbletwo
  \def\@oddhead{}\def\@evenhead{}
  \def\@oddfoot%
   {\rm\hfil\thepage\stepcounter{page}\hskip.5\textwidth\thepage\hfil}
  \let\@evenfoot\@oddfoot
 \else
 \ps@plain
 \fi
}

\documentstyle[hutp]{article}

\makeatletter \input art10.sty \makeatother
\def\starttext{\twocolumn}

\typein{*** or hit [return] for landscape mode(11x8.5) ***}

\newcommand{\beq}{\begin{equation}}
\newcommand{\eeq}{\end{equation}}

\newcommand{\remove}[1]{}
\renewcommand{\theequation}{\thesection.\arabic{equation}}

\newcommand{\lae}{\raisebox{-0.2ex}{$\stackrel{\textstyle<}
{\raisebox{-0.6ex}[0ex][0ex]{$\sim$}}$}}

\newdimen\pmboffset
\def\oldpmb#1{\setbox0=\hbox{#1}%
 \copy0\kern-\wd0
 \kern\pmboffset\raise 1.732\pmboffset\copy0\kern-\wd0
 \kern\pmboffset\box0}

\begin{document}

\title{Oblique Corrections in Technicolor with a Scalar}

\author{
        Christopher D. Carone and Elizabeth H. Simmons \thanks{
        carone@huhepl.harvard.edu, simmons@huhepl.harvard.edu}\\
        Lyman Laboratory of Physics \\
        Harvard University \\
        Cambridge, MA 02138
}
\date{\today}

\renewcommand{\preprintno}{HUTP-92/A022}

\begin{titlepage}

\maketitle

\def\thepage {}        

\begin{abstract}

We study a model in which the electroweak symmetry is dynamically
broken by technicolor interactions and the symmetry-breaking is communicated
to the quarks and leptons by a weak doublet of scalar fields.  The scalars
may be regarded as elementary or as light bound states arising from strong
extended technicolor interactions.  This model is unusual in that it can
provide a heavy top quark without generating large flavor-changing neutral
currents.   Our work focuses on the model's phenomenology, particularly the
oblique electroweak radiative corrections.  We demonstrate that while
oblique corrections and FCNC effects limit the allowed parameter space,
the model remains viable.  In particular, if  the scalar is relatively
heavy, the model can  accommodate a wide range of top  quark masses.

\end{abstract}
\end{titlepage}

\starttext 
\pagestyle{columns} 
\pagenumbering {arabic} 

\section{Introduction} \label {sec:intro}

Most technicolor \cite{e:technicolor} models are are incomplete.  Simple
technicolor, while able to dynamically break the electroweak  symmetry, is
incapable of communicating that symmetry breaking to the quarks and
leptons.  Extended technicolor models \cite{e:etc} can, in principle,
enable the ordinary fermions to acquire mass.  However, an extended
technicolor model which can simultaneously generate a large (of order 100
GeV) top quark mass and avoid large flavor-changing neutral currents
without relying on strong assumptions about the dynamics has
yet to be constructed.  Even forcing the technicolor coupling to `walk'
\cite{e:walk} between the technicolor and extended technicolor scales does
not altogether alleviate this problem.

One technicolor model can be demonstrated to provide both a heavy top
quark and small flavor-changing neutral currents: the technicolor model
containing a weak-doublet of scalars \cite{e:heavy,e:bosonic}.  In this
model, technicolor is solely responsible for electroweak symmetry breaking.
The scalar has a positive mass squared and does not spontaneously break
$SU(2) \times U(1)$; it does, however, acquire a small VEV
through its interactions with the technicolor condensate.  The
scalar's role is the communication of electroweak symmetry breaking to the
ordinary fermions.   Flavor symmetry breaking arises in the Yukawa couplings
of the scalar to fermions, resulting in a standard GIM mechanism.  It is
therefore possible to  produce a heavy top quark without inducing large
strangeness-changing neutral currents \cite{e:heavy}.   Because there is no
particular need to make the technicolor coupling `walk,' even a minimal
(one weak doublet) technifermion sector will suffice.

For those who view a mixture of elementary scalars and dynamical
symmetry breaking with suspicion, it is important to realize
that our model may instead be interpreted as a low-energy effective theory
arising from a model without any elementary scalars. As discussed
in \cite{e:aspects}, an ETC model may be fine tuned so that the scale at
which the ETC interactions would break the technifermion chiral
symmetries lies just below the scale at which the ETC gauge
symmetry is broken. In this case, the ETC interactions do
not directly cause electroweak symmetry breaking but are strong enough to
form a light scalar (fermion--antifermion) bound state. This composite
scalar couples to both ordinary and technifermions and it develops a small
VEV at the technicolor scale.  In other words, it behaves exactly
like the `elementary' scalar described above.  Everything we shall say
about technicolor with a scalar applies equally well to a fine-tuned ETC
model at low energies.

The model's simple structure allows new contributions to various observables
to be calculated directly in terms of the scalar's mass and
couplings.   In the next two sections, we describe the model
in more detail and briefly discuss several points of phenomenology in
order to place a few simple limits on the model parameters.
In section \ref{sec:heavyscalar} we address oblique
electroweak radiative corrections arising in this model when the
scalar's mass exceeds the scale at which the technifermions'
chiral symmetries break; in section \ref{sec:lightscalar} we study oblique
corrections in the opposite limit.  We demonstrate that the
radiative corrections in this model are consistent with current
experimental data for a wide range of top quark mass.
In section \ref{sec:mtmphi} we discuss the experimental limits on the
top quark and scalar masses, and we consider in more detail how these
limits constrain the parameter space in our model.  Our conclusions are
summarized in the final section.

\section{The Model}\label{sec:model}

The gauge boson and fermion content of this model will appear quite
familiar. The gauge group is the direct product of the
technicolor and standard model gauge groups:
$SU(N)_{TC}$ $ \times SU(3)_C\times$ $ SU(2)_W \times
U(1)_Y$. The ordinary, techni-singlet fermions are exactly as
in the standard model.  There are three generations of leptons and quarks in
left-handed doublets and right-handed singlets of $SU(2)$
\begin{eqnarray}
L_{L} & = & {l \choose \nu}_{L} , \ \ \ \ \ l_{R} \nonumber \\
Q_{L} & = & {U \choose D}_{L} ,\ \ \ \ U_{R},\ \ D_{R}
\end{eqnarray}
where $l \equiv (e, \mu, \tau),\ \nu \equiv (\nu_e, \nu_\mu,
\nu_\tau),\ U \equiv (u, c, t),$ and $D \equiv (d, s, b)$.
In the technifermion sector, we will generally assume a minimal
(two techniflavor) model; the analysis is easily extended to
non-minimal models.  Under $SU(2)_W$, the two technifermions
form a left-handed doublet and two right-handed singlets
\begin{equation}
\Upsilon_L = {p \choose m}_L,\ \ \ \ p_R,\ \ m_R
\end{equation}
while their hypercharges $Y(\Upsilon_L)$=0, $Y(p_R)$=${1 \over 2}$,
$Y(m_R)$=$-{1 \over 2}$ are chosen to cancel gauge anomalies.  Both $p$
and $m$ are vectors of $SU(N)_{TC}$.

Dynamical electroweak symmetry breaking occurs as in conventional
technicolor models: when technicolor becomes strong at a scale $4 \pi f$,
the technifermions' chiral symmetries break spontaneously and a condensate
forms,
\begin{equation}
\langle{\bar p p + \bar m m}\rangle \approx 4 \pi f^3,
\end{equation}
where $f$ is the technipion decay constant.  This breaks
$SU(2)_W \times U(1)_Y$ down to $U(1)_{EM}$ and gives
mass to the W and Z bosons.  Also as usual in technicolor models, the
ordinary (techni-singlet) fermions are not given mass directly by the
condensate.  Some additional mechanism is required to
communicate electroweak symmetry breaking to the ordinary fermions.

That mechanism is supplied by a
scalar field $\phi$, an $SU(2)_W$ doublet of mass $M_\phi$.
The scalar's Yukawa couplings to the technifermions:
\begin{equation}
{\cal L}_{\phi T} = \bar\Upsilon_L \tilde\phi\thinspace \lambda_+ p_R\ +\
    \bar\Upsilon_L \phi\thinspace \lambda_- m_R\ +\ \mbox{h.c.}
\label{eeq:ytc}\end{equation}
cause it to acquire an effective vacuum expectation value ($f'$) when the
technifermions condense.  Notice that we are free to redefine
the phases of the right-handed fields in (\ref{eeq:ytc}) so that
$\lambda_+$ and $\lambda_-$ are real.  Using naive dimensional
analysis \cite{e:ndaref} we estimate the vacuum expectation value to be
\begin{equation}
f' \approx {4 \pi \lambda_T f^3 \over M_\phi^2}
\label{eeq:fpapprox}
\end{equation}
where $\lambda_T = {1\over 2}(\lambda_+ + \lambda_-)$ is the scalar's
coupling to the technicondensate.  We reiterate that $\phi$ has
a positive mass squared ($M_\phi^2 > 0$). Unlike the standard model Higgs
doublet, $\phi$ does not cause electroweak symmetry breaking but obtains
a non-zero VEV when technicolor breaks the symmetry.

Because the scalar couples to ordinary fermions as well as technifermions,
\begin{equation}
{\cal L}_{\phi f} = \bar L_{L} \tilde\phi\thinspace \lambda_{l} l_{R}
 + \bar Q_{L} \tilde\phi\thinspace \lambda_{U} U_{R}\ +\
     \bar Q_{L} \phi\thinspace \lambda_{D} D_{R}\ +\ \mbox{h.c.}
\label{eeq:yordf}\end{equation}
the ordinary fermions obtain masses from the diagram sketched in Figure
\ref{fig:scalarvev}
\begin{equation}
m_{f} \approx \lambda_{f} \lambda_T {4 \pi f^3 \over M_\phi^2 }.
\end{equation}
The coupling matrices $\lambda_{f}$ are proportional to the mass
matrices $m_f$ and are the source of flavor symmetry breaking in this
model.  In particular, the quarks' flavor symmetries are broken according
to the same pattern as in the standard model so that the quarks mix via
the usual KM matrix.

This model has four free parameters related to the scalar: the scalar's
mass ($M_\phi$), its couplings to technifermions ($\lambda_+, \lambda_-$),
and its  self-coupling.  The last is of interest only insofar as it affects
the  value of the VEV the scalar acquires from the technicondensate; we
will generally assume it is small enough not to dominate the scalar
potential for the range of scalar masses of interest to us.

All other model parameters can be determined in terms of those four plus
experimental data. We use the well-measured parameters $\alpha$, $G_F$
and $M_Z$ to define the electroweak gauge sector of the model.
Since the scalar coupling to light fermions is small
(because $\lambda_f \propto m_f$) the relationship between $G_F$ and other
standard model parameters is not noticeably altered.  The symmetry-breaking
scale of the technicolor chiral symmetries is defined  by relating $G_F$
to the technipion decay constant $f$, and the induced scalar VEV $f'$
of equation (\ref{eeq:fpapprox}):
\beq
(\sqrt{2} G_F)^{-1} \equiv v^2 =   f^2 + f'^2
\label{eeq:vdef}
\eeq
where $v \approx 250\mbox{ GeV}$.  We define the KM quark-mixing matrix
by using the measured values of the CP-conserving angles \cite{e:pdg}
and using $\epsilon$ to pin down the CP-violating phase $\delta$ as
described in \cite{e:heavy}.  The couplings of the ordinary fermions to
the scalar can be written in terms of their masses:
\begin{equation}
\lambda_{f} \approx {m_{f} M_\phi^2 \over { 4 \pi f^3 \lambda_T}}.
\label{eeq:lamas}
\end{equation}
and those masses are set at their experimental values.  While the top quark
mass is unknown {\it a priori}, the experimental lower bound on $m_t$ is
91 GeV  if the top cannot decay to a charged scalar
\cite{e:cdf} and 55 GeV if it can \cite{e:ua2}.
As we shall find that our model can more readily accommodate a top quark
that does not decay to a charged scalar, we shall generally assume the
top decays only through a W boson.  In section \ref{sec:mtmphi} we will
show how the model's phenomenology is altered if the top can decay
via a charged scalar.

\section{Parameter Space}\label{sec:para}

Of the free parameters, only a subset are relevant for studying any
individual phenomenon, and the most generally useful pair are $M_\phi$
and $\lambda_T$.  Hence, our discussions of the model's phenomenology will be
phrased in terms of determining the allowed area within the
$M_\phi$-$\lambda_T$ plane.  Three simple considerations quickly place
the model within a finite region of this plane.
First, we require the scalar's coupling to the technifermions (and, hence,
the coupling to the condensate) be perturbative:  $\lambda_T < 4 \pi$.
Second, we place a similar requirement on the Yukawa couplings to the
ordinary fermions, the largest of which is the coupling to the top
quark.  Equation (\ref{eeq:lamas}) shows that requiring
$\lambda_f $\lae $4 \pi$ places an upper bound $M_\phi$ for a
given $\lambda_T$.   So the combined limits on $\lambda_T$ and
$\lambda_t$ constrain $M_\phi$ to lie below a value
of order 20 TeV (the precise value depends on the top quark mass $m_t$).
Finding an upper limit on $M_\phi$ in this way is analogous to finding an
upper bound on the scale of the dynamics which gives rise to the large top
quark mass in ETC models.
The third constraint comes from LEP experiments which find no
evidence for a neutral scalar particle of mass less than 48 GeV in the
process $Z \to Z^* \phi$ \cite{e:lepscal}.  As we shall see in section
\ref{sec:mtmphi}, when this is translated into a lower bound on $M_\phi$
there is an approximate cancellation between effects tending to strengthen
and weaken the bound.  In the end, the LEP data require $M_\phi > 46$ GeV,
almost independent of the value of $\lambda_T$.

The allowed region is reduced  when we consider strangeness changing
neutral currents (SCNC), in particular, the measured difference between
the masses of the neutral kaons $K_L$ and $K_S$.  In the standard model, the
largest contribution to the coefficient of the $\Delta S = 2$ operator
$(\bar s \gamma^\mu (1 + \gamma_5) d)^2$ comes from the box diagram of
figure \ref{fig:boxdia}a.  This contribution is of order
\begin{equation}
{g^4 (\xi^K_c)^2 m_c^2 \over { 16 \pi^2 M_W^4}}
\label{eeq:stddels}
\end{equation}
where $\xi^K_q \equiv V_{qd} V^*_{qs}$ and the factor of $m_c^2$ is
required because of GIM cancellation among box diagrams with different
internal quark flavors\footnote{The graph with internal top quarks
contributes less because $\xi^K_t$ is very small.}.
Equation (\ref{eeq:stddels}) appears to account quite well for the
observed mass difference.  In our model, there are additional contributions
to the strangeness-changing Hamiltonian from box diagrams with
internal scalars (figure \ref{fig:boxdia}b,c).  Providing that
$M_\phi \geq m_t$, the dominant pieces of diagrams \ref{fig:boxdia}b,c are,
respectively,
\begin{equation}
{g^2 (\xi^K_t)^2 m_t^4 \over {16 \pi^2 M_W^2 M_\phi^2}} \left( {M_\phi^2
\over {4 \pi f^3 \lambda_T}}\right)^2
\ \ \ {\rm and}\ \ \
{(\xi^K_t)^2 m_t^4 \over {16 \pi^2 M_\phi^2}} \left( {M_\phi^2
\over {4 \pi f^3 \lambda_T}}\right)^4
\label{eeq:newhdels}
\end{equation}
Here, the powers of top quark mass arise from a combination of GIM and the
scalar-quark coupling constants.  The factor of $m_t^4$ overcomes the small
mixing angles $\xi^K_t$.    In order for the new contributions
not to render the predicted mass difference inconsistent with experiment,
we require that (\ref{eeq:stddels}) exceed (\ref{eeq:newhdels}).
This sets\footnote{A more detailed discussion of
the calculation was presented in \cite{e:heavy}.}
an upper bound on $M_\phi$ for given $\lambda_T$ as shown in figure
\ref{fig:roughps}.

We shall refer to the strangeness-changing neutral current
bound on $M_\phi$ and $\lambda_T$ as the `SCNC-line' for brevity.
This bound places an increasingly tight upper limit on  $M_\phi$ as
$m_t$ is increased; for the smallest allowed $m_t$ of
91 GeV,  one finds $M_\phi < 8.9$ TeV.  Even for this smallest value of
$m_t$, the SCNC bound supersedes the  `consistency' requirement $\lambda_t
< 4\pi$ by restricting $\lambda_t$ to values smaller than $\sim 3.6$.  As
discussed in \cite{e:heavy}, it also excludes the region of parameter space
in which $\epsilon'/\epsilon$ is predicted to be much smaller than the
measured value.

The area of interest to us in the $M_\phi$-$\lambda_T$ plane is shown in
figure \ref{fig:roughps}, with the SCNC-line and the perturbative
constraints  drawn in explicitly.   For now, the left-hand edge of the
graph  (at $M_\phi$ = 30 GeV) will approximate the third boundary of the
allowed triangle; that boundary will be made explicit in section
\ref{sec:mtmphi}.  Also shown in figure \ref{fig:roughps} is the curve
(obtained from equations (\ref{eeq:fpapprox}) and (\ref{eeq:vdef}))  along
which $M_\phi$ and the technicolor scale $4\pi f$ are equal. This
separates two physically different regions of parameter space which we
will explore separately in the following sections.  We say that the scalar
is `heavy' compared to the technicolor scale when its mass is at  least a
few TeV.

\section {Oblique Corrections for $M_\phi > \Lambda$}
\label {sec:heavyscalar}

Oblique \cite{e:oblique} radiative corrections  (those affecting gauge
boson vacuum polarizations rather than vertices)  are powerful sources of
information about new physics
\cite{e:holtern,e:golran,e:pestak,e:takeuchi,e:kenlyn,e:kenlang}.  We begin
this section by reviewing the sources of oblique corrections in this model.
We then examine the oblique corrections generated by the scalar when $M_\phi$
lies above the technicolor scale $4 \pi f$.  To do this we will move from the
complete high-energy theory through an effective theory at scales
$4\pi f < \mu < M_\phi$ where the scalar has been integrated out to the
effective theory relevant at the low energies where radiative corrections
are measured.

We adopt the parametrization of \cite{e:pestak} which identifies
three interesting combinations of vacuum polarizations
\begin{eqnarray}
S& = & - 16 \pi \left[{d\over dq^2} \Pi_{3B}\right]_{q^2 = 0} \\
T& = & {4 \pi \over s^2 c^2 M_Z^2} \left[\Pi_{11} -
\Pi_{33}\right]_{q^2 = 0} \\
U& = & 16 \pi \left[ {d\over dq^2} (\Pi_{11} - \Pi_{33}) \right]
_{q^2 = 0}
\end{eqnarray}
where $s$ and $c$ are the sine and cosine of the weak mixing angle,
respectively.  Because of the approximate custodial symmetry in our model,
$U$ will be small compared to $T$ \cite{e:pestak}; we will neglect $U$
from here on.  Then $S$ ($T$) may be identified as measuring the
discrepancy between predictions of our model and those of the minimal
standard model in the weak isospin conserving (violating) sector.  Because
radiative corrections in the standard model depend on the top and higgs
masses, one must choose an $m_t$ and $m_H$ at which to define $S = T = 0$.
Following \cite{e:pestak,e:takeuchi} for later convenience, we choose the
reference masses to be $m_t = 150$ GeV and $m_H = 1$ TeV.

There are several contributions to the $S$ parameter in
our model.  The largest comes from the technicolor sector;  as estimated in
\cite{e:pestak}, this is  approximately consistent with the simple formula
\begin{equation}
S_{0} \approx .3 \, {N_{TC}\over 3} \, {v^2\over f^2}
\label{eq:ptsest}
\end{equation}
In the technicolor models considered in \cite{e:pestak},
the technipion decay constant $f$ is lower than the weak scale $v$ only if
more than one weak doublet of technifermions exists.  However, in our
model, the small VEV acquired by the scalar also pulls $f$ lower than $v$
(c.f. equation \ref{eeq:vdef}).
Therefore, for a fixed number of technicolors and technifermion doublets,
$S_0$ will be larger in our model than in the
technicolor theories studied in \cite{e:pestak}.  A second contribution to
$S$ comes from the heavy fermion doublet $(t,b)$:
\begin{eqnarray}
S_{tb} &\approx& -{1\over 6\pi}\log({m_t^2\over m_{t,{\rm ref}}^2}).
\end{eqnarray}
This is identical to the heavy top-quark contribution
in the standard model, and and is small compared to $S_0$.
The four-technifermion operators
formed when the scalar is integrated out of the theory can make yet
another  contribution to $S$; we will find that this is also suppressed
relative to $S_0$.

The primary contributions to $T$ come from the
scalar's coupling to fermions.  The Yukawa terms in the Lagrangian
(\ref{eeq:ytc}, \ref{eeq:yordf}) possess an $SU(2)_L \times SU(2)_R$ global
symmetry so long as $\lambda_u = \lambda_d$ and  $\lambda_+ =
\lambda_-$.   When the couplings are unequal, the $SU(2)_R$ symmetry is
broken, generating a contribution to the $T$ parameter.  Symmetry
considerations show that the leading dependence of $T$ on
$(\lambda_u - \lambda_d)$ and $(\lambda_+ - \lambda_-)$ is at least
quadratic.  We will word the argument in terms of $(\lambda_+ -
\lambda_-)$ for brevity, but it applies equally well to $(\lambda_u -
\lambda_d)$.  We start with the observations that
the weak bosons form a {\bf3} of weak isospin and that the weak boson two-point
function $\Pi_{aa}$ is symmetric under interchange of the two bosons.
Those facts imply that $\Pi_{aa}$ transforms according to the
symmetric part of the weak isospin representation ({\bf3}$\times${\bf3}),
\beq ({\rm{\bf 3}}\times{\rm{\bf 3}})_{\rm s} = {\rm{\bf 1} }_{\rm s} +
{\rm {\bf 5} } _{\rm s} \label{eq:c3c3}
\eeq
Therefore $T$, which parametrizes weak isospin violation, must transform as
a {\bf 5} of weak isospin.  To see what this implies for the Yukawa terms'
contributions to $T$, we can rewrite (\ref{eeq:ytc}) in the form
\beq
\bar\Upsilon_L \Phi \lambda \Upsilon_R
\label{eeq:twoterm}
\eeq
where $\Phi = (\tilde\phi \phi)$ transforms as a ({\bf 2},{\bf 2}) under
$SU(2)_L\times SU(2)_R$ and $\Upsilon_R = (p,m)_R$ transforms as a ({\bf
1},{\bf 2}).  Treating the coupling matrix  $\lambda =
\mbox{diag}(\lambda_+,\lambda_-)$ as a spurion, we see that $\lambda$ must
transform as a ({\bf 1},{\bf 1})+({\bf 1},{\bf 3}); the $SU(2)_R$ singlet
is proportional to the sum of the Yukawa couplings and the triplet is
proportional to their difference.  Two factors of $\lambda$ are required to
produce anything transforming as a  ({\bf 1},{\bf 5}).   This means that the
leading contribution to $T$ from the technicolor sector will depend
quadratically on $(\lambda_+ - \lambda _-)$.  Any remaining
dependence on $\lambda_+$ and $\lambda_-$ must be of the form $(\lambda_+ +
\lambda_-)^N$.

In the ordinary fermion sector, the most serious
breaking of $SU(2)_R$ comes from the fact $\lambda_t >> \lambda_b$ as
evidenced by the disparity in the top and bottom quark masses.
This contributes to $T$ exactly as in the standard model \cite{e:pestak}:
\beq
T_{tb}
\approx \frac{3}{16 \pi s^2 c^2 M_Z^2}(m_t^2 - m_{t,{\rm ref}}^2).
\label{eeq:stdto}
\eeq
As expected from the symmetry argument
and the smallness of $\lambda_b$, this is quadratic in $m_t$.

Since the technifermions condense at energies well above those where
oblique corrections are measured, we will study the phenomenology
of our model by constructing a low energy effective field theory.
In order to quantify the effect of $\lambda_+ \neq \lambda_-$ on $T$,
we will first integrate out the heavy scalar, and then map the resulting
technifermion operators onto an effective chiral lagrangian of technipion
fields.  At energy scales below $M_\phi$, the scalar-technifermion
couplings (\ref{eeq:ytc}) give rise to two kinds of four-technifermion
operators suppressed by the square of the scalar mass
\begin{eqnarray}
{\lambda_+ \lambda_- \over M_\phi^2} \bar\Upsilon^i_L
\Upsilon^m_R \bar\Upsilon^j_L \Upsilon^n_R \epsilon_{ij} \epsilon_{mn} \\
{1\over M_\phi^2} \bar\Upsilon^i_L \Upsilon^m_R
\left( \begin{array}{cc} \lambda_+^2&0\\ 0& \lambda_-^2 \end{array}
\right)
\bar\Upsilon^m_R  \Upsilon^i_L
\label{eeq:fiop}\end{eqnarray}
where $i,j,m,n$ are $SU(2)_{L,R}$ indices.  The first kind of operator is
a singlet under both $SU(2)$'s.  The second is an $SU(2)_L$ singlet (since
the scalar respects the electroweak symmetry) but has both singlet and
triplet pieces under $SU(2)_R$:
\begin{equation}
{(\lambda_+^2 + \lambda_-^2)\over 2 M_\phi^2} \bar\Upsilon_L \Upsilon_R
\bar\Upsilon_R \Upsilon_L \ +\
{(\lambda_+^2 - \lambda_-^2)\over 2 M_\phi^2} \bar\Upsilon_L \Upsilon_R
{\bf \tau}_3 \bar\Upsilon_R \Upsilon_L .
\label{eeq:hiop}
\end{equation}
The triplet piece of (\ref{eeq:hiop}), proportional to $\lambda_+ -
\lambda_-$, contributes to $T$.  This is analogous to certain
$SU(2)_R$-violating four-technifermion operators arising in ETC models
when the ETC bosons are integrated out as discussed in \cite{e:appel}.

The next step is to see how the four-technifermion operator in
(\ref{eeq:hiop}) contributes to an effective chiral lagrangian of
technipions.  We use the  conventional language classifying chiral
lagrangian operators by their transformation properties under
global $SU(2)_L \times SU(2)_R$, where
$SU(2)_L$ is identified with $SU(2)_W$ and the $T_{3}$ component of
$SU(2)_R$ is identified  with $U(1)_Y$.  The kinetic energy terms for the
electroweak gauge bosons and the (eaten) technipions are
\begin{equation}
{\cal L}_0 = -{1\over 2}{\rm Tr}\left[W^{\mu\nu} W_{\mu\nu}\right]
-{1\over 2}{\rm Tr}\left[B^{\mu\nu} B_{\mu\nu}\right] +
{v^2\over 4} {\rm Tr}\left[D^\mu \Sigma^\dagger D_\mu \Sigma\right] .
\end{equation}
Here $W^\mu = {e\over 2 s} W^\mu_a \tau_a$, $B^\mu = {e\over 2 c} B^\mu
\tau^3$ and $\Sigma = \exp(2 i \pi/f)$ is a non-linear representation of the
technipion fields transforming as $\Sigma \to L \Sigma R^\dagger$
under $SU(2)_L \times SU(2)_R$.  The covariant derivative is then
defined as $D^\mu \Sigma = \partial ^\mu \Sigma - i W^\mu \Sigma +
i \Sigma B^\mu$. To determine how the effects of (\ref{eeq:hiop}) feed
down into the chiral lagrangian, we will treat the symmetry-breaking
parameter ${1\over2}(\lambda_+^2 - \lambda_-^2){\bf \tau}_3$ as a spurion
field in the adjoint representation of $SU(2)_R$.  We may then
construct chiral lagrangian operators which contain the spurion field
and contribute to $T$.  The leading operators will be of lowest order
in the momentum expansion and will contain the fewest powers of the spurion.

Upon constructing chiral lagrangian operators involving the spurion,
we find that there is only one  operator at leading order
\footnote{The operator ${1\over2}(\lambda_+^2 - \lambda_-^2)^2 v^2 {\rm Tr}
[{{\bf \tau}_3\over 2} \Sigma D^\mu \Sigma^\dagger
{{\bf \tau}_3\over 2} \Sigma D_\mu \Sigma^\dagger ]$,
which is of the same order, differs from (\ref{eeq:toper}) only by an
$SU(2)_L$ invariant.} that contributes to $T$
\begin{equation}
\xi '\, {v^2 (\lambda_+^2 - \lambda_-^2)^2\over 2}
\left[{\rm Tr} {\tau_{3}\over 2} \Sigma D^\mu \Sigma^\dagger \right]^2 .
\label{eeq:toper}
\end{equation}
\begin{equation}
\delta_T \equiv {\lambda_+ - \lambda_- \over {\lambda_+ + \lambda_-}}
\end{equation}
The unknown coefficient $\xi '$ is of order one, by naive dimensional
analysis, but its sign is not known {\it a priori}; we will set
$\xi ' = \pm 1$.  Then, this operator's contribution to $\Pi_{33}(0)$
is $\pm (\lambda_+^2 - \lambda_-^2)^2 v^2/4$, so that an amount
\begin{equation}
\Delta T \approx \pm  {(\lambda_+^2 - \lambda_-^2)^2 \over 2}
\left( {{\rm TeV} \over M_\phi} \right)^4 = \pm 8 \, \delta_T^2
\left({\lambda_T \over M_\phi}\right)^4.
\label{eeq:scalt}
\end{equation}
is added to $T$. The effect of the
four-technifermion operators on the $T$ parameter is, thus,
quadratic\footnote{Because our four-fermion operators
arise from scalar exchange, the operators found in \cite{e:appel} to give
an effect linear in the fermion mass splitting do not arise in our model.}
in $\delta_T$ as expected by our symmetry arguments and is of unknown sign.

For completeness we should also comment on the contributions to $S$ made by
the four-technifermion operators (\ref{eeq:fiop}).  It was noted earlier that
the leading contribution to $S$ in our model comes from the technicolor
sector.  In the chiral lagrangian language, the leading operator
contributing to $S$ is independent of the scalar coupling to
technifermions:
\begin{equation}
{\xi\over 16\pi^2}{\rm Tr}\left[\Sigma^\dagger W^{\mu\nu}
\Sigma B_{\mu\nu}\right] .
\label{eeq:firsts}
\end{equation}
Here naive dimensional analysis \cite{e:ndaref} leads us to expect that
$\xi$ is of order one.  We make the identification
\beq
S_0 = \frac{1}{\pi}\,\xi
\eeq
where $S_0$ is given by eq. (\ref{eq:ptsest}).  While the four-technifermion
operators (\ref{eeq:fiop}) will give rise to additional chiral lagrangian
operators contributing to $S$, those additional operators will be
suppressed relative to (\ref{eeq:firsts}) by a factor of at least
$(\lambda_T f /M_\phi)^4$.  Their contributions to $S$ may therefore
be neglected.

We are now ready to compare our model's predictions for $S$ and $T$ with
a global fit to current low-energy and LEP data recently performed by
Peskin and Takeuchi \cite{e:pestak,e:takeuchi}.
\footnote{We thank T. Takeuchi for providing us with his fitting
program and data.} The results of their well-known analysis are shown
in figure \ref{fig:stplotone}.  Also shown in that figure are our model's
predictions in the absence of $SU(2)_R$ violation in the scalar-technifermion
couplings ($\delta_T = 0$).  We see that when $\delta_T = 0$, electroweak
radiative corrections directly set limits on  the top quark
mass. For $N=2$ technicolor, the range of top quark  masses allowed
at 95\% confidence level is roughly    125 GeV $< m_t <$ 200 GeV;
for $N=4$ technicolor, the range is  150 GeV $< m_t <$ 205 GeV.
The standard model predictions for $S$ and $T$  are shown in
figure \ref{fig:stplotone} for comparison; the range of allowed top
quark masses is similar.

If the splitting $\delta_T$ is non-zero, the radiative corrections are
somewhat different.  While the contribution of $\delta_T$ to the $S$
parameter is negligible (as discussed earlier), the contribution to $T$ can
be large and is of unknown sign.  Hence, as we turn on $\delta_T$, the
range of allowed top quark masses will shift.  For any given $m_t$ to be
consistent with the data, it is necessary for $\Delta T$ to fall within
certain bounds.  We plot the allowed range of $\Delta T$ as a function of
$m_t$ in  Figure \ref{fig:trange}.  Note that if the sign of $\Delta T$
were known, we would, instead, be able to place a direct upper or lower
bound on $m_t$ within this model.

To assess the degree to which the data on oblique corrections restricts the
model, we shall return to the $M_\phi$-$\lambda_T$ plane.
The analysis of \cite{e:pestak} found that for models in which one  expects
$S > 0$ (e.g. technicolor models) the data constrain $S$ to lie below 0.9
at 95\% c.l.  When we apply this constraint to $S$ in our model (in fact,
requiring $S_0 < 0.9$) we obtain a lower bound on $M_\phi$ as a function of
$\lambda_T$ as shown in figure \ref{fig:fallsshort}.  This slightly
reduces the allowed portion of the heavy-scalar region.  Turning to the $T$
parameter, we recall that a light top quark is consistent with the oblique
corrections data only if $\Delta T$ is sufficiently large (and
positive).  From equation (\ref{eeq:scalt}) and the fact that
$\delta_T$ is defined to be $\leq 1$,
it is clear that obtaining a large value of $\Delta T$ depends
on having a large enough ratio $\lambda_T/M_\phi$.  If the model is to
accommodate the lightest top quark (91 GeV) currently allowed by
experiment \cite{e:cdf}, then the minimum value of $\lambda_T/M_\phi$
required is about 0.7.  As indicated in figure \ref{fig:fallsshort} the
SCNC-line has already excluded the portions of heavy-scalar parameter space
where $\lambda_T/M_\phi$ falls below that value.

To summarize: when the scalar is heavier than the technicolor scale, the
model can potentially accommodate any top quark mass between 90 and
$250^+$ GeV without running afoul of the data on oblique radiative corrections.
This is in distinct contrast to the ETC models considered in
\cite{e:pestak,e:takeuchi} in which only relatively light top quark masses
were allowed.  In addition, we have seen that the leading contribution of
$\phi$ to $T$ is quadratic in $\delta_T$  rather than linear as in the
leading contribution from ETC-induced four-technifermion operators
explored in \cite{e:appel}.

\section {Oblique Corrections for $M_\phi \ll \Lambda$}
\label {sec:lightscalar}

We will now study electroweak radiative corrections in a technicolor model
with a light scalar doublet, that is, for $M_\phi \ll \Lambda$ .   In this
limit, it is no longer sensible to integrate out the scalar fields, and the
four light degrees of freedom in $\Phi$ will remain explicitly in the
low-energy spectrum.  Our approach in this section will be to construct a
gauged chiral Lagrangian that includes the scalar fields, and to
estimate the contributions to $S$ and $T$ from the vacuum polarization
diagrams containing light physical scalars at one-loop.

To begin, we must identify of the light physical scalars
appearing in the low energy spectrum.  As in
section \ref{sec:heavyscalar}, we rewrite the scalar doublet in
the matrix form
\beq
\Phi = \left[ \begin{array}{cc} \overline{\phi^0} & \phi ^+ \\
                                - \phi ^-         &  \phi ^0
              \end{array} \right]
\eeq
and adopt the non-linear representation $\Sigma$ for the technipion
fields. The kinetic terms for the scalar fields can then be written
\beq
{\cal L}_{K.E.} = \frac{1}{2} \mbox{Tr}(D_\mu \Phi ^\dagger
D ^\mu \Phi) + \frac{f^2}{4} \mbox{Tr}( D_\mu \Sigma ^\dagger
D^\mu \Sigma)
\label{eq:cke}
\eeq
Because eq. (\ref{eq:cke}) contains couplings of the form
\beq
\begin{array}{lcr}
W_\mu \partial ^\mu \pi & \mbox{ and } & W_\mu \partial ^\mu \Phi
\end{array}
\label{eq:ccouple}
\eeq
it follows that some combination of the technipions and the light scalars
becomes the longitudinal component of each weak gauge boson.  To
determine which combination is absorbed, we will use the fact that
$\Phi^\dagger \Phi \propto 1$ to rewrite $\Phi$ in terms of an isosinglet
scalar field $\sigma$, and a unitary matrix $\Sigma '$ :
\beq
\Phi \equiv \frac{(\sigma + f')}{\sqrt{2}} \, \Sigma '
\label{eq:csp}
\eeq
where
\beq
\Sigma ' = \exp (\frac{2 \, i \, \pi '}{f'})
\eeq
$\Sigma '$ is defined in analogy to $\Sigma$, and contains an isotriplet of
pions, $\pi '$, that mix with our original technipions.  Note that $\sigma$
has been shifted by $f'$ in (\ref{eq:csp}) to assure the appropriate
normalization of the $\Sigma '$ kinetic term.  The kinetic energy
terms (\ref{eq:cke}) now become
\beq
{\cal L}_{K.E.} = \frac{1}{2} \partial_\mu \sigma \partial ^\mu \sigma
+\frac{f^2}{4} \mbox{Tr} (D_\mu \Sigma ^\dagger D^\mu \Sigma)
+\frac{(\sigma+f')^2}{4} \mbox{Tr} (D_\mu {\Sigma '} ^\dagger D^\mu \Sigma
')
\label{eq:cnke}
\eeq
By examining the W-pion couplings in (\ref{eq:cnke}), we may
identify the linear combination of $\pi$ and $\pi '$  that is absorbed
by the gauge bosons.  The orthogonal linear combination
corresponds to a triplet of physical pions.  We find
\beq
\pi _a  =  \frac{f \pi + f' \pi '}{\sqrt{f^2 + {f'} ^2}}
\eeq
and
\beq
\pi _p  = \frac{-f' \pi + f \pi '}{\sqrt{f^2 + {f'} ^2}}
\eeq
where $\pi_a$ and $\pi_p$ are the absorbed and the physical pions,
respectively.  It is also clear from (\ref{eq:cnke}) that we obtain
the correct gauge boson masses providing that
\beq
f^2 +{f'}^2 = \frac{s^2 c^2}{\pi \alpha} M_Z^2
\equiv v^2
\label{eq:cmwcon}
\eeq
Thus, we recover eq. (\ref{eeq:vdef}).  In our treatment of electroweak
radiative corrections, we will work in unitary gauge, where the spectrum
includes $\pi_p$, $\sigma$ and the (massive) gauge bosons only.

Now let us consider the scalar potential.  The leading terms are
\beq
V(\Phi) = \frac{M_\phi^2}{2} \mbox{Tr} (\Phi ^\dagger \Phi) + \kappa
\left[ \mbox{Tr} (\Phi ^\dagger \Phi) \right] ^2
\eeq
After applying the redefinition (\ref{eq:csp}) we obtain
\beq
V(\sigma) =  \frac{M_\phi^2}{2} (\sigma  + f')^2  + \kappa (\sigma + f')^4
\label{eq:cpot}
\eeq
and hence the mass of the isosinglet $\sigma$ is given by
\beq
M_\sigma^2 = M_\phi^2 + 12 \kappa {f'}^2
\label{eq:crhomass}
\eeq

The leading potential (\ref{eq:cpot}) is altered by the interactions
of the scalar with the technipions.  To write down all such interactions
consistent with the chiral symmetry, we recall
(c.f. eq.(\ref{eeq:twoterm})) that the coupling of the scalar to the
techniquark doublet $\Upsilon$ can be written
\beq
\Upsilon _L
\left( \begin{array}{cc} \overline{\phi^0} & \phi ^+ \\
                                - \phi ^-         &  \phi ^0
              \end{array} \right)
\left( \begin{array}{cc} \lambda _+ & 0 \\
                           0       & \lambda _-
       \end{array} \right)
\Upsilon _R \equiv \Upsilon _L \Phi \lambda \Upsilon _R
\eeq
where $\lambda$ is the matrix of Yukawa couplings shown above.
In the spirit of a spurion analysis, we pretend that the matrix $\Phi
\lambda$ transforms according to
\beq
(\Phi \lambda) \rightarrow L \, (\Phi \lambda) \, R^\dagger
\eeq
and build all possible invariants.  The simplest term we can construct is
\beq
c_1 \cdot 4 \pi f^3 \, \mbox{Tr}(\Phi \lambda \Sigma ^\dagger) +
\mbox{h.c.}
\label{eq:cc1}
\eeq
where we have used naive dimensional analysis \cite{e:ndaref}
to rescale the coefficient $c_1$ so that it is of order unity. We shall
continue this practice  henceforth.  Eq. (\ref{eq:cc1}) contributes a
piece linear in $\sigma$ to the scalar potential,
\beq
V_{\lambda}(\sigma) = -
c_1 \cdot 4 \sqrt{2} \pi f^3 (\lambda_+ + \lambda_-) \sigma
\label{eq:clinear}
\eeq
This linear term appears to give $\sigma$ a non-vanishing vacuum expectation
value.   However,  eq. (\ref{eq:cmwcon}) alone is not sufficient to fix
both $f$ and $f'$, so we have the freedom to require the linear terms
in $V(\sigma)$ to vanish
\beq
f' M_\phi^2 - 4 \kappa {f'}^3 - c_1 \cdot 4 \sqrt{2} \pi f^3
(\lambda_+ + \lambda_-) = 0
\label{eq:ccon}
\eeq
which implies that in the small $\kappa$ limit
\beq
f' \approx \sqrt{2} c_1 \frac{(\lambda_+ + \lambda_-)(4 \pi
f^3)}{M_\phi^2}
\label{eq:csmallk}
\eeq
For simplicity, we will restrict ourselves to the case where $\kappa$ is
negligible, so that (\ref{eq:crhomass}) reduces to
$M_\sigma \approx M_\phi$, and eq. (\ref{eq:csmallk}) is also valid.
Then eqs. (\ref{eq:csmallk}) and (\ref{eq:cmwcon}) completely specify
both decay constants in terms of other model parameters.

While we have determined that $\sigma$ has mass $M_\phi$ at lowest
order, we have not yet commented on the mass of the pion multiplet.
The physical pions receive a common mass at ${\cal O}(\lambda)$ in the
chiral expansion, from the piece of (\ref{eq:cc1}) quadratic in $\pi_p$.
One finds
\beq
M^2_{\pi} = c_1 \cdot \sqrt{2} \frac{4 \pi f}{f'} v^2
(\lambda _+ + \lambda _-) \approx \frac{v^2}{f^2} \, M_\phi^2
\label{eq:ccommass}
\eeq
where we have assumed $c_1 > 0$.  These masses are split at ${\cal
O}(\lambda^2)$ by the effects of the term
\beq
c_2 \cdot f^2
\mbox{Tr}(\Phi\lambda\Sigma^\dagger\Phi\lambda\Sigma^\dagger) +
\mbox{h.c.}
\label{eq:cc2}
\eeq
Evaluating the quadratic terms in eq. (\ref{eq:cc2}) one obtains
\beq
-c_2 \cdot v^2 \left[ \frac{1}{2}(\lambda^2_+ +
\lambda^2_-){\pi^0_p}^2 + 2 \lambda_+\lambda_- \pi^+_p \pi^-_p \right]
\eeq
from which we conclude that the splitting between the masses of $\pi^+_p$
and $\pi^0_p$ is
\beq
m_0^2 - m_+^2 = c_2 \cdot v^2 (\lambda_+ - \lambda_-)^2
\label{eq:cdmsq}
\eeq
where $c_2$ can be of either sign.  Eq. (\ref{eq:cc2}) also contains
an ${\cal O}(\lambda ^2)$ contribution to the common pion mass which we
ignore in comparison to (\ref{eq:ccommass}).

The pion mass splittings violate the custodial symmetry and therefore
contribute to the $T$ parameter through the one-loop diagrams shown
in Figure \ref{fig:vacpol}.  Evaluating the diagrams in  Figure
\ref{fig:vacpol}a, we find
\[
i g^{\mu\nu} (\Pi _{11} - \Pi_ {33})  + (k^\mu k^\nu \mbox{ terms} )  =
\mbox{    }
\]
\beq
\frac{1}{4} \int \frac{d^4 p}{(2\pi)^4} \left\{
\frac{(k+2p)^\mu (k+2p)^\nu}{\left[ (k+p)^2 - m_+^2\right] }
\left[ \frac{1}{p^2-m_0^2} - \frac{1}{p^2-m_+^2} \right] \right\}
\label{eq:cint}
\eeq
which vanishes when $m_0 = m_+$, as required.  We
evaluate these integrals using dimensional regularization in $4+\epsilon$
dimensions, and obtain the following
\beq
\Pi_{11} - \Pi_{33} = \frac{1}{64\pi^2} (m_+^2 - m_0^2)\left[
\left(\frac{2}{\epsilon}+\gamma _E - \ln (4\pi) \right) +
\ln(\frac{m_0^2}{\mu ^2}) +\frac{1}{2} \right] + \cdots
\label{eq:cevalint}
\eeq
where the dots refer to terms of ${\cal O} ((m_+^2-m_0^2)^2)$ and higher.
If we work in the $\overline{\mbox{MS}}$ prescription, where the first term
in square brackets is cancelled by a counterterm, we find that the leading
correction to $T$ is given by
\beq
\Delta T = \frac{1}{16 \pi s^2 M_W^2} (m_+^2 - m_0^2) \ln
(\frac{m_0^2}{\mu^2})
\label{eq:cdt}
\eeq
assuming that the logarithmic term is dominant.  Combining this result with
(\ref{eq:cdmsq}), we finally obtain
\beq
\Delta T = c_2 \cdot\frac{\pi}{s^2 M_W^2} v^2
\left(\frac{\lambda _+ - \lambda _-}{4 \pi} \right)^2 \ln
(\frac{M_\pi ^2}{\Lambda ^2})
\label{eq:ctest}
\eeq
where we have replaced $m_0 \approx M_\pi$ in the logarithm,
and where we have set $\mu$ equal to the chiral symmetry breaking scale
$\Lambda \approx 4\pi f$.

The Feynman diagrams that each contain a single internal $\sigma$ line
can be evaluated using the same approach outlined above.  Unlike the
diagrams that only have triplet pions running around the loops, the
diagrams in Fig. \ref{fig:vacpol}b. yield contributions to $T$ of the form
$\ln(M_\phi^2/\Lambda^2)$, as well as $\ln(M_\pi^2/\Lambda^2)$.
If we again assume that the logarithmically enhanced terms dominate, we
may incorporate their contribution to $T$ by rewriting
(\ref{eq:ctest}) as
\beq
\Delta T = c_2 \cdot\frac{\pi}{s^2 M_W^2} v^2
\left(\frac{\lambda _+ - \lambda _-}{4 \pi} \right)^2 \left[ A \ln
(\frac{M_\pi ^2}{\Lambda ^2}) + B \ln(\frac{M_\phi ^2}{\Lambda ^2}) \right]
\label{eq:cntest}
\eeq
where $A$ and $B$ are given by
\beq
A = 1 - \frac{f^2}{v^2}\frac{(M_\pi^4 - 2 M_\pi^2 M_\phi^2)}
{(M_\pi^2 - M_\phi^2)^2}
\label{eq:cA}
\eeq
\beq
B= - \frac{f^2}{v^2}\frac{M_\phi^4}{(M_\pi^2 - M_\phi^2)^2}
\label{eq:cB}
\eeq
Eq. (\ref{eq:cntest}) includes the logarithmic contributions from
all the diagrams in Fig.\ref{fig:vacpol}, to lowest order in the
squared-mass difference $m_+^2 - m_0^2$.

The assumption of logarithmic enhancement in the preceding analysis is
valid providing that we restrict ourselves to the region of parameter space in
which the technipions are light.  Chiral perturbation theory itself is only
sensible if $M_\pi \ll 4\pi f$, otherwise we would not be justified in
treating the technipions as pseudogoldstone bosons.  To be safe, we
will restrict ourselves to the region of the $M_\phi$-$\lambda_T$ plane
in which $M_\pi < \pi f$.  In analogy to QCD, we expect chiral perturbation
theory to be valid in this region and, in addition, the logarithms
will be no smaller than $\sim 2.8$.

Notice that in our estimate of $\Delta T$ in (\ref{eq:cntest}), the isospin
violation originated solely from the splitting of the technipion masses.
We must also consider the one-loop contributions to $T$ that are
nonvanishing when the technipion triplet is taken to be degenerate.
For contributions of this type, the spurion $\lambda$ enters the
vacuum polarization diagrams through isospin violating $W$-$\pi$-$\pi$
couplings.  While the gauge boson - technipion couplings from the kinetic
terms (\ref{eq:cnke}) are isospin conserving, we may obtain isospin
violating couplings at higher order in the chiral expansion.
At ${\cal O}(\lambda)$, there are exactly two terms containing
$W$-$\pi$-$\pi$ couplings, namely
\[
d_1 \cdot\frac{f}{4\pi} \mbox{Tr}(D_\mu \Phi \lambda D^\mu \Sigma
^\dagger) + \mbox{h.c.}
\]
\beq
d_2 \cdot \frac{f}{4\pi} \mbox{Tr}(\Phi\lambda \Sigma ^\dagger
D^\mu \Sigma D_\mu \Sigma ^\dagger) + \mbox{h.c.}
\label{eq:cd1d2}
\eeq
Although the spurion $\lambda$ transforms as a {\bf 1} + {\bf 3} under
$SU(2)_R$, only the singlet contributes to the trilinear couplings, when
$d_1$ and $d_2$ are chosen to be real.  For complex coefficients,
(\ref{eq:cd1d2}) yields $W$-$\pi$-$\pi$ couplings that transform
as a {\bf 3} under $SU(2)_R$, but they are CP violating.  As we have seen
in section \ref{sec:model}, $\lambda_+$ and $\lambda_-$ can be made
real without introducing unremovable phases, so the technicolor sector of our
model is CP conserving, and we conclude that the vertices
described above are not present.  We will not encounter $W$-$\pi$-$\pi$
couplings that transform nontrivially until ${\cal O}(\lambda ^2)$, through
terms of the form\footnote{These terms directly split the $W_{1}$ and
$W_{3}$ masses, but this contribution to $T$ is smaller than the pion
mass splitting effects by the large logarithmic factors in (\ref{eq:cntest}) }
\beq
\frac{\beta}{16\pi^2} \mbox{Tr}(\Phi\lambda D_\mu \Sigma^\dagger
\Phi\lambda D^\mu \Sigma^\dagger) + \mbox{h.c.}
\label{eq:cvertex}
\eeq
Notice that this term transforms as a {\bf 1} + {\bf 3} + {\bf 5}
under $SU(2)_R$.  The one-loop vacuum polarization diagrams that
incorporate a single $W$-$\pi$-$\pi$ vertex from (\ref{eq:cvertex}), and
an isospin conserving vertex from (\ref{eq:cnke}) has a component that
transforms as a {\bf 5}, and therefore can contribute to the $T$ parameter
by the arguments given in section \ref{sec:heavyscalar}.  Dimensional analysis
tells us that the results are of the order
\beq
\Pi_{11} - \Pi_{33} \propto \frac{M^2}{64\pi^2} \left(\frac{\lambda_+-
\lambda_-}{4\pi}\right)^2
\ln (\frac{M^2}{\Lambda^2})
\label{eq:civv}
\eeq
where we have used dimensional regularization to evaluate the loop, and
where $M$ represents either $M_\pi$ or $M_\phi$.  However, when we
considered the pion mass splitting, we obtained
\beq
\Pi_{11}-\Pi_{33} \propto
\frac{(4\pi v)^2}{64 \pi^2}
\left(\frac{\lambda_+-\lambda_-}{4\pi} \right)^2
\ln (\frac{M^2}{\Lambda^2})
\label{eq:civms}
\eeq
Clearly (\ref{eq:civms}) dominates over the (\ref{eq:civv}) by a factor
of $(4\pi v)^2/ M^2$, which is large given that $4\pi v > \Lambda \gg M$.
We conclude that the leading correction to $T$ from the technicolor
sector in our model comes from the mass splitting of the physical pion
triplet, as given by equation (\ref{eq:cntest}), with subleading
effects suppressed by powers of $M^2 / (4\pi v)^2 $

We also need to estimate the size of $S$ for the technicolor model with a
light scalar.  The tree-level technicolor contribution from the operator
\beq
\frac{\xi}{16 \pi^2}\mbox{Tr}(\Sigma^\dagger W^{\mu\nu}\Sigma
B_{\mu\nu} )
\eeq
may be identified with $S_0$ of eq. (\ref{eq:ptsest}) as before; for
$N_{TC} = 2$, we obtain
\beq
S_0 \approx 0.2 \, \frac{v^2}{f^2}.
\label{eq:csnought}
\eeq
There are also contributions to $\Pi_{3B}$ from the  one-loop
diagrams with internal scalars shown in Fig.\ref{fig:pi3b}. The leading
log effects are found by a straightforward analysis to be
\beq
\Delta S = \frac{1}{12\pi}
\frac{{f'}^2}{v^2} \ln \left( \frac{M_\pi^2}{\Lambda^2}\right)
\label{eq:cs}
\eeq
Notice that as $M_\phi \rightarrow \infty$ one finds $\Delta S \rightarrow
0$ so that $S$ matches onto the value found for technicolor with a heavy
scalar in section {\ref{sec:heavyscalar}.

We may now use the current experimental limits on $S$ and $T$ to restrict
the allowed region of the $M_\phi$-$\lambda_T$ plane.  Starting with $S$,
we require
\beq
S_0 + \Delta S < 0.9
\eeq
for our model to be consistent with experiment \cite{e:takeuchi}.
This restriction confines us to a strip of parameter space
bounded from above by the $S=0.9$ line, and below by the SCNC line, as
shown in Fig.\ref{fig:sline}.  Since it is conceivable that the experimental
limits on $S$ may be revised in time, we varied our constraint on $S$ by
$\pm 10$\%, and found that the allowed area shown in Fig.\ref{fig:sline} was
not appreciably altered.  In performing the numerical analysis, all unknown
coefficients of order unity were set identically equal to $1$.

Turning to constraints imposed by $T$, we must now combine the
contributions $T_{tb}$ and $\Delta T$.  The first of these is
the same as in the standard model and was given in eq. (\ref{eeq:stdto}).
The maximal leading log contribution to $\Delta T$ (obtained from eq.
(\ref{eq:cntest}) with $\delta_T = 1$) is shown in Table I for a sample
of points in the allowed region.  Comparison with eq. (\ref{eeq:stdto})
shows that  $\Delta T$ is much smaller than $T_{tb}$.  Hence, even for maximum
$\delta_T$,  the restriction $T$ places on $m_t$ in our model is very
similar to the restriction $T_{tb}$ places on $m_t$ in the standard
model.  Referring to  Fig. \ref{fig:stplotone}, we see that without
additional contributions to  $T$ of positive sign, the lightest top quark
consistent with the experimental  constraints on $T$ ranges from 100 to
125 GeV, for $S$ ranging from 0 to 0.2.  This new lower bound on $m_t$
supersedes the bound $m_t > 91$ GeV used to generate the SCNC
limits.  The effect is to shift the SCNC line  upwards by a small amount
in  the $M_\phi$-$\lambda_T$ plane, very slightly reducing the allowed
region of parameter space we  isolated in Fig.\ref{fig:sline}.

\begin{table}
\begin{center}
\begin{tabular}{c||cccc}
&\multicolumn{4}{c}{$M_\phi$ (GeV)} \\
$\lambda_T$ & 50 & 100 & 150 & 200 \\ \hline
0.005 & $\pm$0.0003 & - & - & - \\
0.007 & $\pm$0.0007 & - & - & - \\
0.01  & $\pm$0.0014 & - & - & - \\
0.02  &      -      & $\pm$0.0041 & $\pm$0.0027 & - \\
0.04  &      -      & $\pm$0.0161 & $\pm$0.0135 & $\pm$0.0099 \\
0.06  &      -      &     -       & $\pm$0.0303 & $\pm$0.0253 \\
0.08  &      -      &     -       & $\pm$0.0519 & $\pm$0.0459 \\
0.1   &      -      &     -       &       -     & $\pm$0.0707 \\
0.12  &      -      &     -       &       -     & $\pm$0.0990 \\
\end{tabular}
\end{center}
\caption{$\Delta T _{\mbox{max}}$}
\end{table}

\section{Experimental Lower bounds on $M_\phi$ }
\label{sec:mtmphi}

To complete our phenomenological discussion, we consider experimental
lower bounds on $M_\phi$. The first of these derives from searches for
new scalar particles and therefore applies whether or not the top quark can
decay to the light charged scalar in our model.  The others depend on the
available top quark decay modes, so the cases $m_t < m_+$ and $m_t > m_+$
will be discussed separately.

Recent LEP searches for a neutral
higgs boson, $H$, in the process ($Z \to Z^* H$) set a limit
$M_H > 48$ GeV \cite{e:lepscal}.  Two steps are involved in translating this
bound into a limit on our parameter $M_\phi$.  First, we recall that the
Z-Z-scalar coupling is proportional to the scalar VEV in both our model and
the standard model. This reduces the  expected $\sigma$ production rate in
our model by a factor of $(f'/v)^2$ relative to the standard model rate of
Higgs production.  In consequence, the lower bound on $M_\phi$ is pulled
below $48$ GeV where $f \sim v$.   Second, we note that $M_\sigma$ differs
slightly from $M_\phi$ when the  higher order corrections from
(\ref{eq:cc2}) are included \beq
M_\sigma^2 = M_\phi^2 - 2 (\lambda_+^2 + \lambda_-^2) f^2 .
\eeq
This tends to push the lower bound on $M_\phi$ above 48 GeV as
$\lambda_T$ grows larger. When combined, these competing effects largely
cancel.  The net result is that the LEP data imply a lower bound on
$M_\phi$ of $\sim 46$ GeV with only limited dependence on $\lambda_T$ as
illustrated in Figure \ref{fig:sline}.   Experimental lower bounds on the
mass of light  charged scalars also exist  \cite{e:lepscal}, but these
limits are numerically  weaker while our charged scalars are expected to be
at least  as heavy as  our neutral scalar.

If the top quark
decays almost exclusively through a W boson, experiment tells us that
$m_t > 91$ GeV \cite{e:cdf}.  In this case,  the physical charged scalar
must weigh more than 86 GeV.  The relationship (\ref{eq:ccommass}) between
$M_\pi$ and $M_\phi$ allows us to translate  this
into a lower bound on $M_\phi$.  As shown in figure \ref{fig:sline},
this bound partially supersedes the lower limit on $M_\phi$ which we just
derived from $Z \to Z^* \phi$.  We see from figure \ref{fig:sline}
that almost all of the parameter space previously found to be ``allowed''
in our model is consistent with a top quark which is unable to decay to
the physical charged scalar.

Now let us examine the possibility that the top quark {\it is} able to decay
to the physical charged scalar.  Since our study of the oblique corrections
indicates that the top quark is unlikely to be heavier than a few hundred
GeV, $M_\phi$ will be similarly restricted and much of the previously
``allowed'' parameter space will be excluded.  Further, experimental lower
bounds on the top quark mass will provide upper, rather than lower,
limits on $M_\phi$. Data on $\Gamma_W$ from CERN and FNAL sets a lower
bound $m_t > 55$ GeV which is independent of the top quark's decay modes
\cite{e:ua2}; the charged scalar must be still lighter.
Direct searches for top quarks decaying through charged scalars
have excluded \cite{e:dirser} a small amount of additional parameter space
for relatively light $m_t$ and $M_\phi$.

If the top quark is heavier than the charged scalar, the leading
contribution to the box diagrams (\ref{fig:boxdia}b,c) responsible
for SCNC in our model is now of order
\begin{equation}
{g^2 (\xi^K_t)^2 m_t^2 \over {16 \pi^2 M_W^2}} \left( {M_\phi^2
\over {4 \pi f^3 \lambda_T}}\right)^2
\ \ \ {\rm and}\ \ \
{(\xi^K_t)^2 m_t^2 \over {16 \pi^2}} \left( {M_\phi^2
\over {4 \pi f^3 \lambda_T}}\right)^4
\label{eeq:newldels}
\end{equation}
rather than as in (\ref{eeq:newhdels}). For very light $M_\phi$, the top
quark can be as light as 55 GeV \cite{e:ua2}, and the SCNC-line lies at
relatively low $\lambda_T$.  As $M_\phi$ is increased, the effective lower
limit on $m_t$ must also increase so that the top quark will still be
kinematically allowed to decay to a charged scalar (recall from
(\ref{eq:ccommass}) that $M_\phi \approx m_+ (f/v)$).  Since the new
contributions (\ref{eeq:newldels}) to SCNC in our model grow with the top
quark mass, this pushes the SCNC-line towards higher $\lambda_T$.  A
comparison of figures \ref{fig:sline} and \ref{fig:newsline}
illustrates the steeper dependance of the SCNC-line on $M_\phi$ in the
case where top quark decay to a charged scalar is allowed.

\section {Conclusions}\label {sec:conc}

We have estimated the electroweak radiative corrections in
a technicolor model with a scalar doublet.  This model provides a
simple mechanism for generating fermion masses while
avoiding large flavor changing neutral currents;
this distinguishes it from most extended technicolor schemes.

We have considered two distinct regions of this model's parameter space.
In the region where the scalar was heavy compared to the technicolor scale
($M_\phi > 4 \pi f$) we integrated out the scalar and wrote down an
effective theory with higher dimension operators suppressed by powers of
$M_\phi$.  We then considered the effects of the leading operators on
the model's phenomenology.  Where the scalar was light and the scalar
coupling to technifermions (or, equivalently, to the condensate) was
relatively small, the technifermions did not receive large dynamical masses
from their coupling to the scalar and it was consistent to treat the
$\bar T T$ bound states as goldstone bosons.  We studied this region by
constructing a chiral expansion in the two small parameters
$M_\phi / 4 \pi f$ and $\lambda_T / 4 \pi$.   In both regions, we found
a significant area of parameter space in which the model is consistent
with current experimental limits on the $S$ and $T$ parameters,
for a wide range of top quark masses.  This is in contrast to the
situation for the  extended technicolor models considered in
\cite{e:pestak,e:takeuchi} which are estimated to have such large
contributions to $T$ that all  but relatively light top quark masses are
excluded.

While we have not explicitly examined the piece of allowed parameter space
lying  between the heavy-scalar and light-scalar regions, the reader will
have noticed that our phenomenological bounds have been extended through
this region as well.  In the intermediate region, the scalar mass $M_\phi$ is
on the order of or a bit less than the technicolor scale $4 \pi f$.  Hence when
technicolor condenses, the scalar is still in the theory and acquires a VEV,
allowing it to give mass both to the ordinary fermions and to the
technifermions.  Unlike the case in the light scalar region, however, the
technifermion bound states are no longer light, and therefore cannot be
adequately described as goldstone bosons, as in section \ref{sec:lightscalar}.
The correct Lagrangian describing the physics in this region of parameter
space is that of a chiral technifermion model incorporating both massive
technifermions and the (eaten) goldstone bosons of chiral symmetry
breaking.  However, the phenomenological bounds we placed on the
heavy-scalar region using naive dimensional analysis will hold equally well
in this region since the symmetry-breaking parameter $\lambda f /
M_\phi$ remains small \cite{e:ndaref}.

We hope that the current work will generate further interest in the
viability of this model.

\vspace{0.5in}

\centerline{\bf Acknowledgments}
We thank H. Georgi and R.S. Chivukula for helpful conversations, and C.
Jessop and M. Franklin for discussions of top quark searches. E.H.S.
acknowledges the financial support of a Superconducting Super Collider
fellowship from the Texas National Research Laboratory  Commission  (TNRLC)
and the hospitality of the Aspen Center for Physics where some of the
writing was completed. {\it This work was  supported in part by the National
Science Foundation under grant PHY-87- 14654, and the Texas National
Research Laboratory Commission, under grant  RGFY9106.}



\begin{figure}[p]
\caption{Fermion-technifermion interaction through scalar exchange.
This diagram is responsible for fermion mass generation
after the technifermions condense.}
\label{fig:scalarvev}
\end{figure}

\begin{figure}[p]
\caption{$\Delta S =2$ box diagrams: (a) is leading diagram in standard
model; (b),(c) are leading diagrams with extra scalars in loop.}
\label{fig:boxdia}
\end{figure}

\begin{figure}[p]
\caption{Preliminary limits on the allowed region of the
$M_\phi$-$\lambda_T$ plane. The allowed region lies within the solid
lines, i.e., above the SCNC line and below the $\lambda_T = 4\pi$
line.}
\label{fig:roughps}
\end{figure}

\begin{figure}[p]
\caption{Predictions for $S$ and $T$ in our model, with $\delta_T = 0$.
The $N_{TC} = 2$ and $N_{TC} = 4$ results are shown, in comparison to
the predictions of the Standard Model with a 1 TeV Higgs.  The points
plotted show the variation in $T$ for a range in top quark mass  from 75
to 250 GeV.  The experimental 68\% and 95\% confidence contours
are also shown.}
\label{fig:stplotone}
\end{figure}

\begin{figure}[p]
\caption{Allowed range in $\Delta T$ as a function of $m_t$.}
\label{fig:trange}
\end{figure}

\begin{figure}[p]
\caption{Allowed region for heavy scalar, including limits from oblique
corrections.  The line $\lambda_T/M_\phi = .7$ is shown to lie below the
SCNC line for a 91 GeV top quark; thus the model can accommodate a top
quark
this light.}
\label{fig:fallsshort}
\end{figure}

\begin{figure}[p]
\caption{Vacuum polarization diagrams contributing to
$\Pi_{11}-\Pi_{33}$}
\label{fig:vacpol}
\end{figure}

\begin{figure}[p]
\caption{Vacuum polarization diagrams contributing to $\Pi_{3B}$}
\label{fig:pi3b}
\end{figure}

\begin{figure}[p]
\caption{Allowed region of the $M_\phi$-$\lambda_T$ plane, showing
the $S=0.9$ line, and LEP limits on light neutral scalar mass. The
top-quark is assumed not to decay into the light scalar.}
\label{fig:sline}
\end{figure}

\begin{figure}[p]
\caption{Allowed region of the $M_\phi$-$\lambda_T$ plane, showing
the $S=0.9$ line, and LEP limits on light neutral scalar mass. Here, the
top-quark can decay into the light scalar.}
\label{fig:newsline}
\end{figure}

\end{document}